\documentclass{pasj00}
\begin{document}
\SetRunningHead{R. Nagino and K. Matsushita}{Suzaku observation of S0
galaxy NGC 4382}

\Received{2009/11/04}
\Accepted{2010/04/14}

\title{The Abundance Pattern of O, Ne, Mg, and Fe in the Interstellar Medium
of S0 Galaxy NGC 4382 Observed with Suzaku}

\author{Ryo \textsc{Nagino}
   and Kyoko \textsc{Matsushita}}
\affil{Tokyo University of Science, 1-3 Kagurazaka, Shinjyuku-ku, Tokyo,
  Japan, 162-8601}
\email{nagino@rs.kagu.tus.ac.jp; matusita@rs.kagu.tus.ac.jp}


\maketitle

\begin{abstract}
We derived O, Ne, and Mg abundances in the interstellar medium (ISM) of a
 relatively isolated S0 galaxy, NGC 4382, observed with the Suzaku XIS
 instruments and compared the O/Ne/Mg/Fe abundance pattern to those
 of the ISM in elliptical galaxies. The derived temperature and Fe
 abundance in the ISM are about 0.3 keV and 0.6--2.9 solar,
 respectively. The abundance ratios are derived with a better accuracy than
the  abundances themselves:  O/Fe, Ne/Fe, and Mg/Fe ratios are 0.3, 0.7,
 and 0.6, respectively, in solar units. The O/Fe ratio is
 smaller than that of the ISM in elliptical galaxies, NGC 720, NGC 1399,
 NGC 1404, and NGC 4636, observed with Suzaku. Since O, Ne, and Mg are predominantly
 synthesized by supernovae (SNe) of type II, the observed abundance
 pattern indicates that the contribution of SN Ia products is higher in
 the S0 galaxy than in the elliptical galaxies.
 Since the hot ISM in early-type galaxies is an accumulation of stellar mass
 and SN Ia products, the low O/Fe ratio in the ISM of NGC 4382 
 reflects a higher rate of present SNe Ia, or stars containing more SN Ia
 products than those in elliptical galaxies.
\end{abstract}

\section{Introduction}

Early-type galaxies have a hot, X-ray emitting interstellar medium
(ISM), which is considered to be gravitationally confined (e.g.,
\cite{Form1985,Math2003}).  X-ray observations of the metal abundances in the
ISM of early-type galaxies provide a key to understanding the history of
star formation and the evolution of galaxies, because the metals in the
ISM come from type Ia supernovae (SNe Ia) and stellar mass loss. 

A lot of  observational evidence suggests that 
a significant fraction of present
early-type galaxies have transformed from late-type galaxies
(e.g., \cite{Dres1997,Fasa2000,Treu2003,Post2005,Smith2005}).
From z$\sim$0.5,
the fraction of S0 galaxies  in clusters of galaxies has increased,
whereas, that of spirals has decreased, with no evolution in the fraction of ellipticals
\citep{Koda2004,Desai2007}.
\citet{Pogg2009} found that these changes appear more strongly in
less massive clusters with lower velocity dispersion.
These results suggest that spiral galaxies changed into S0
galaxies at z$<$0.5, falling into clusters of galaxies.
The  metallicity  of stars in galaxies reflects in the star formation
history, therefore, it is an important parameter for understanding the
evolution of galaxies.

Optical observations indicate that  Mg/Fe and $\alpha$/Fe ratios of stars are super-solar in
the cores of bright early-type galaxies and increases with galactic mass
(e.g., \cite{Fabe1992,Wort1992,Nela2005,Thom2005,Grav2007}).
This overabundance of Mg relative to Fe is the key indicator that galaxy
formation occurred before a substantial number of SNe Ia could explode
and contribute to lowering the these ratios
(e.g., \cite{Bern2003,Nela2005,Smith2006,Grav2007,Pipi2009}).
However, absorption-line indices that account for
abundance ratios also depend on the age distribution of
stars.  Optical spectroscopy is limited within the very center of galaxies.

Using X-ray observations, we can directly determine the metal abundances
of the ISM, and  constrain the stellar metallicity of the entire galaxy.
The atomic data for lines at X-ray wavelengths are  simpler than
for those in optical spectra, and the structure of the hot ISM is also much
simpler than stellar population data.
Therefore, we can estimate the temperature and metallicity of the hot ISM through X-ray
spectra with small systematic uncertainties.
XMM-Newton EPIC and RGS provided the means to measure O and Mg abundances in some
systems, but reliable results have been obtained only for several central
galaxies in groups and clusters (e.g., \cite{Xu2002,Matsu2003,Tamu2003,Matsu2007a}).
The ISM in such regions might be polluted by the gas of clusters or
groups (e.g., \cite{Matsu2001,Matsu2002,Nagi2009}). 

With the Suzaku X-ray satellite, O, Ne, and Mg abundances of four elliptical
galaxies, NGC 720 \citep{Tawara2008}, NGC 1399 and
NGC 1404 \citep{Matsu2007b}, and NGC 4636 \citep{Haya2009}, have been measured. 
The XIS detector onboard Suzaku \citep{Mitsu2007} can constrain
 O, Ne, and Mg abundances well because its energy resolution is better, and
its background is lower than any previous X-ray CCD detector \citep{Koya2007}.
According to the new solar abundance
table of \citet{Lodd2003}, the abundance ratios of these elliptical
galaxies are close to the solar values. 
Assuming the SN II abundance pattern of \citet{Iwa1999}, about 80\% of
 Fe is synthesized by SNe Ia \citep{Matsu2007b}. 

The abundance pattern of the hot ISM in S0 galaxies 
observed with Suzaku has not been reported.
In this paper, we present
 the temperature and abundances of O, Ne, Mg, and Fe in the ISM within
 4~$r_e$ of the S0 galaxy
NGC 4382, observed with Suzaku.
 Here, $r_e$ is the effective
radius of the galaxy. 
For NGC 4382, $r_e$ corresponds to 0.91 arcmin \citep{deVau1991}.
The distance to NGC 4382 is 16.8 Mpc \citep{Tully1988}, and its redshift,
z=0.002432, is taken from the NASA/IPAC extragalactic database (NED).
NGC 4382 is located in the outskirts of the Virgo cluster, 1.7 Mpc from
cD galaxy M87. 
The central stellar velocity dispersion of the galaxy, 179 km
s$^{-1}$ and its $[\alpha/{\rm Fe}]$ value, $0.12\pm0.06$ \citep{McD2006},
are much smaller than those of the four elliptical galaxies,
observed with Suzaku.
These values were derived only within two arcsec of the center of NGC
4382, and we do not know the metallicity of stars outside of the central region.
So far, NGC 4382 has been observed with ROSAT \citep{Fabb1994}, ASCA
\citep{Kim1996}, Chandra \citep{Siva2003,Athey2007} and XMM-Newton
\citep{San2006,Nagi2009}. These X-ray observations suggest that the
temperature of the ISM in NGC 4382 is about 0.3--0.4 keV, which is much lower than
those of the four early-type galaxies observed with Suzaku.
Since no the intracluster medium (ICM) around NGC 4382 has been detected,
this galaxy is suitable for
investigating heavy elements in the ISM of the galaxy itself.

Throughout this paper, we use a Hubble constant of $H_0 = 70$
${\rm km}$ ${\rm s^{-1}}$ ${\rm Mpc^{-1}}$. 
We adopt the new solar abundance table of \citet{Lodd2003}. 
The abundances of O and Fe have increased by about 70\% and 60\%, respectively.
Unless otherwise specified, errors are quoted at 90\% confidence.

\section{Observation and data reduction}

\begin{figure}
  \begin{center}
    \FigureFile(80mm,80mm){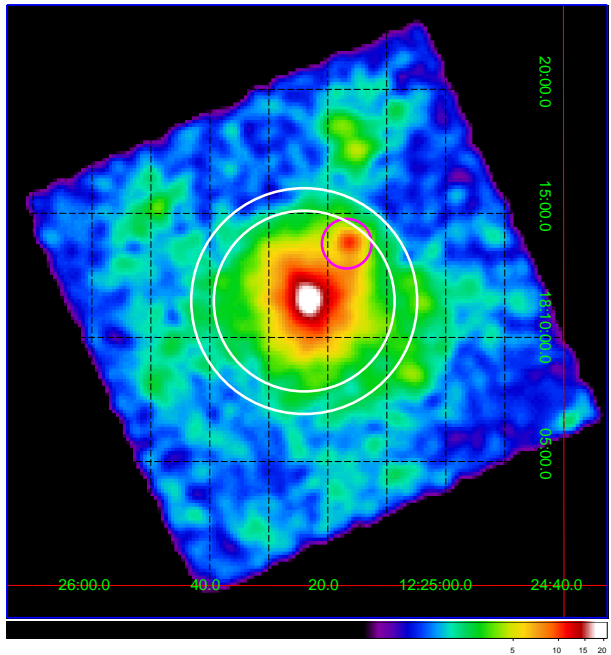}
  \end{center}
  \caption{Raw XIS image of NGC 4382 within an energy range of 0.3--2.0
 keV. XIS0, 1, and 3 images were added in sky coordinates. Non-X-ray
 background (NXB)
 and cosmic X-ray background (CXB) were not subtracted. Image
 was smoothed with a Gaussian of $\sigma=0.56$ arcmin.
 White circles represent regions within 4~$r_e$ and 5~$r_e$, respectively. 
Point source indicated by magenta circle is
 excluded from spectra.}\label{fig:ximg}
\end{figure}

Suzaku observed  NGC 4382 in June 2008 with
an exposure of 99 ksec in the XIS nominal position, pointing at
the center of NGC 4382.
Figure \ref{fig:ximg} shows a 0.3--2.0 keV image of NGC 4382. 
We analyzed only the XIS data in this paper, although Suzaku
observations were performed with both XIS and HXD. The XIS instrument
consists of three sets of X-ray CCDs (XIS0, 1, and 3). XIS1 is a
back-illuminated (BI) sensor, while XIS0 and 3 are front-illuminated
(FI) ones. In this observation, these instruments were operated in the
normal clocking mode (8 s exposure per frame), with the standard
$5\times5$ or $3\times3$ editing mode.

Our analysis was performed using XSPEC version 11.3.2ag and HEAsoft version
6.6.2.
We processed the XIS data with {\tt xispi}, using the calibration files distributed on
04-03-2009. Then, we cleaned the XIS data by assuming thresholds 
of  $>5{\degree}$ on the
Earth elevation angle,  and  $>20{\degree}$ on the day-earth elevation angle.
The total exposure time for both FI and BI sensors amounts to $\sim$99 ksec after standard data
screening. Event screening with cut-off rigidity (COR) was not performed
on our data.

XIS response files were calculated using the {\tt xisrmfgen}
response matrix file (RMF) generator \citep{Ishi2007}, version
02-28-2009. Ancillary response files (ARF) were calculated using
{\tt xissimarfgen}, version 01-08-2009, assuming flat-sky emission.
We note that  ARF files assuming a point-source, a $\beta$-model profile, or
flat-sky are almost the same within the energy range of 0.4 to 5 keV, 
except for normalization.
Slight degradation of the energy resolution was considered in the RMF,
and a decrease in the low-energy transmission of the XIS optical blocking
filter (OBF) was included in the ARFs.
To subtract the contribution of non-X-ray background (NXB), we
used {\tt xisnxbgen}, which generates the NXB spectrum from a dark
Earth database \citep{Tawa2008}. 

\section{Spectral analysis}

As shown in Figure \ref{fig:ximg}, we extracted the spectra within a 4~$r_e$
 circle centered on the optical center of NGC 4382 from NED. 
We also used the region outside 5~$r_e$ to estimate the background.
Northwest of NGC 4382, a point-like X-ray source was detected at
0.3--2.0keV. We excluded this source from each spectrum.

We used the spectra from the  BI (XIS1) and FI (XIS0 and 3) sensors for the energy
ranges of 0.4--5.0 keV and
0.6--5.0 keV, respectively, since
subtracting background lines is difficult above 5.0 keV,
and the ISM does not emit photons above 5.0 keV. The energy
range around the Si K-edge (1.82--1.84 keV) was ignored as a result of  a problem
in the response matrix.
The remaining total photon counts with XIS0, 1, and 3 are
3490, 6604, and 3617 counts, respectively.
Although two calibration sources emitting strong lines above 5 keV
are located at the corners of each XIS,
spectral analysis is not affected  below 5 keV.
Therefore, we included  these regions, and analyzed the spectra below 5 keV.
The spectra of
XIS detectors (XIS0, 1, and 3) were fitted simultaneously with a model
consisting of the ISM emission, unresolved  discrete sources, cosmic
X-ray background (CXB), and Galactic components. Here, the spectra of
the NXB were subtracted.

\subsection{The ISM component}

We adopted a single-temperature vAPEC model \citep{Smith2001} as the ISM
thin thermal emission component (hereafter, the 1T model).
The ISM in early-type galaxies originates mainly from
stellar mass loss, and most of the metals in these stars are the
products of SNe.
The metal abundances of He, C, N, and Al were fixed at  solar values.
We organized the abundances of other heavy elements into five groups: O;
Ne; Mg; the Si group (Si, S, Ar, and Ca); and the Fe group (Fe and Ni),
and allowed them to vary.
This is because  significant detections of the ionized O, Ne, Mg, Si, and  Fe lines
are expected.
In this study, we assume
that  abundances of S, Ar, and Ca are the same as that of Si, taking into account the
metal synthesis process in SNe.
These lines were not  significantly observed, and hence we could not
determine  the abundance of each line.
In our Galaxy,  Ni/Fe ratios synthesized from both SN Ia and SN II are solar values  
\citep{Edva1993}.

This component was subjected to a common absorption with a
fixed column density ($N_{\rm H}$) at the Galactic value of \citet{Dick1990},
$2.5\times10^{20} {\rm cm}^{-2}$.

\subsection{Unresolved discrete sources}
\label{sec:us}

For unresolved point sources,
we added a power-law
model, in  which the photon index was fixed at 1.6 
(\cite{Irwin2003}).
The power-law model
 described the total spectra of discrete sources in early-type galaxies well, 
and was successfully used for this role in previous studies
\citep{Blan2001,Rand2004}.
This component, as well as the ISM component, was also subjected to a common absorption with fixed
$N_{\rm H}$ at the Galactic value of \citet{Dick1990}.

\subsection{CXB and Galactic emission}

For the CXB component, we adopted a power-law model
with a photon index $\Gamma = 1.4$ \citep{Kushi2002}.
 Galactic emission arises mainly  from the local hot bubble (LHB) and
the Milky Way halo (MWH). We used a two-temperature APEC model for
these components and fixed the temperatures of the LHB and MWH emission
at 0.1 and 0.3 keV, respectively, following previous studies
(e.g., \cite{Komi2009,Kona2009,Haya2009,Yama2009}).
 CXB and MWH components were subjected to absorption at the
Galactic value, while the LHB was not.

\begin{figure}
  \begin{center}
    \FigureFile(80mm,80mm){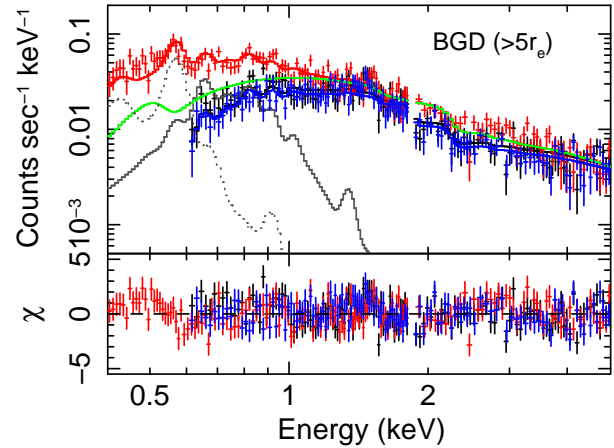}
  \end{center}
  \caption{Spectra outside of 5~$r_e$ observed
 with XIS0 (black), XIS1 (red), and XIS3 (blue). 
 These spectra are fitted with a sum of the CXB and Galactic models.
 Lower panel shows fit residuals. Here, for simplicity,
 only model components for the XIS1 spectra are shown. Green and gray lines
 correspond to CXB and Galactic emission components, respectively.}\label{fig:spec_bgd}
\end{figure}

As shown in Figure \ref{fig:ximg}, we fitted the spectra with the NXB subtracted
outside of 5~$r_e$ from the center of NGC 4382
to estimate the contribution from the CXB and Galactic emission.
In this study,
 we assumed that there is no emission from the galaxy beyond 5~$r_e$.
The normalizations of  CXB and Galactic components were allowed to vary.
Figure \ref{fig:spec_bgd} shows the
best-fit model and spectra for the CXB and Galactic components. The spectra
are well described by the model with a reduced-$\chi^2$ value of 1.23:
there were 419 degrees of freedom.

\subsection{Spectral fitting within 4~$r_e$}

\begin{figure*}
  \begin{center}
    \FigureFile(160mm,80mm){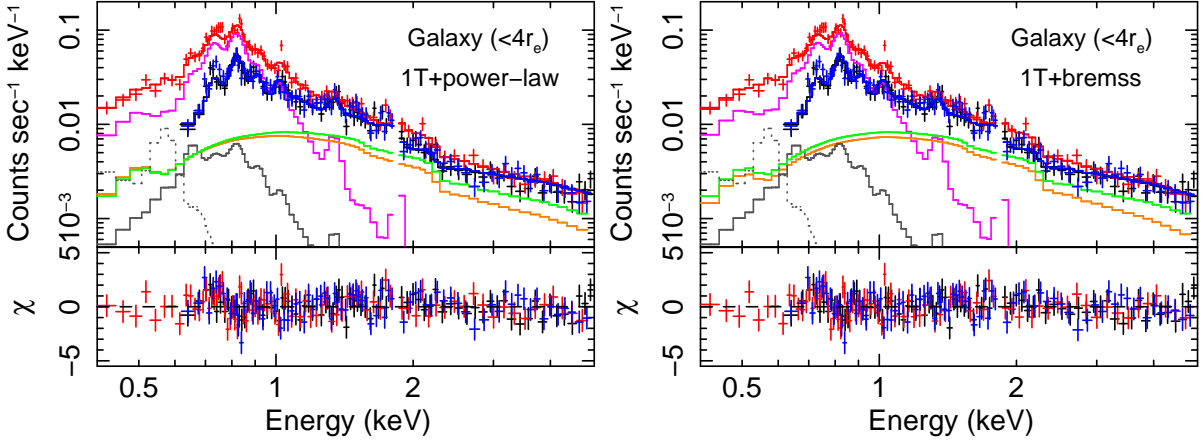}
  \end{center}
 \caption{Spectra within 4~$r_e$ observed
 with XIS0 (black), XIS1 (red), and XIS3 (blue). 
 These spectra are fitted with 1T+power-law (left), and 1T+bremss (right) models.
 Lower panels show fit residuals.
 For simplicity, only model components for XIS1 spectra are
 shown. Magenta and orange lines represent ISM components and unresolved sources, respectively.
 Green and gray lines correspond to CXB and Galactic emission components, respectively.}\label{fig:spec}
\end{figure*}

Next, we fitted the spectra within 4~$r_e$ and outside of 5~$r_e$
of XIS0, 1, and 3 simultaneously with the 1T
model for the ISM, two power-law models for unresolved sources and  CXB, and
two APEC models for Galactic emission. 
Here, we assumed that the normalizations of the ISM and unresolved source
components were fixed at 0 beyond 5~$r_e$.
We also assumed that the CXB and Galactic emission within
4~$r_e$ and beyond 5~$r_e$ have the same surface brightness and
spectral parameters.
Figure \ref{fig:spec} shows the XIS spectra fitted in this way. The derived $\chi^2$,
ISM temperature, and abundances are summarized in Table \ref{tab:fit}.
This fit provides a rather small reduced-$\chi^2$ of $\sim1.15$,
although residual structures appear around 0.8 keV.

\begin{table*}
\begin{center}
\caption{Results of spectral fittings within 4~$r_e$ of NGC 4382 with 1T models}
\label{tab:fit}
\begin{tabular}{lcccccc}
\hline \hline
model & kT   & O       & Ne & Mg & Fe & $\chi^2$/d.o.f. \\
      & (keV) & (solar) & (solar) & (solar) & (solar) &                 \\ \hline
1T+power-law & $ 0.311 _{- 0.019 }^{+ 0.024 } $ & $  0.33 _{-  0.16 }^{+  0.67 } $ & $  0.70 _{-  0.31 }^{+  1.36 } $ & $  0.63 _{-  0.33 }^{+  1.29 } $ & $  1.02 _{-  0.43 }^{+  1.91 } $ &   759 /   658 \\
1T+bremss & $ 0.308 _{- 0.018 }^{+ 0.023 } $ & $  0.30 _{-  0.13 }^{+  0.47 } $ & $  0.64 _{-  0.26 }^{+  0.95 } $ & $  0.56 _{-  0.28 }^{+  0.90 } $ & $  0.93 _{-  0.37 }^{+  1.34 } $ &   754 /   658 \\
\hline
\end{tabular}
\end{center}
\begin{center}
\begin{tabular}{lccc}
\hline \hline
model & O/Fe & Ne/Fe & Mg/Fe \\
      & (solar) & (solar) & (solar) \\\hline
1T+power-law & $  0.33 _{-  0.11 }^{+  0.19 } $ & $  0.69 _{-  0.13 }^{+  0.15 } $ & $  0.62 _{-  0.27 }^{+  0.27 } $ \\
1T+bremss & $  0.32 _{-  0.10 }^{+  0.18 } $ & $  0.68 _{-  0.13 }^{+  0.15 } $ & $  0.60 _{-  0.27 }^{+  0.27 } $ \\
\hline
\end{tabular}
\end{center}
\end{table*}

\section{Results}

\subsection{ISM temperature and abundances}

Table \ref{tab:fit} summarizes the best-fit parameters of temperature
and abundances. The ISM temperature within 4~$r_e$ is about 0.3 keV.
This temperature is lower than those of  early-type galaxies,
 NGC 720, NGC 1399, NGC 1404, and NGC 4636, which have been observed
with Suzaku (\cite{Matsu2007b,Tawara2008,Haya2009}).
This temperature is consistent with those derived from previous observations with ROSAT \citep{Fabb1994},
Chandra \citep{Siva2003,Athey2007}, and XMM-Newton \citep{Nagi2009}.
On the other hand, \citet{San2006} detected a higher ISM
temperature of about 0.4 keV with XMM-Newton EPIC. 

The abundances are less constrained, and Fe abundance within 4~$r_e$
is $\sim0.6$--$2.9$ solar with large errors. 
This value is consistent with those of the  early-type galaxies,
NGC 720, NGC 1399, NGC 1404, and NGC 4636. The other abundances, O, Ne, and
Mg, are mostly consistent with the Fe abundance within large error bars,
but tend to have lower values than that of  Fe.

\subsection{Abundance ratios}

\begin{figure*}
  \begin{center}
    \FigureFile(80mm,80mm){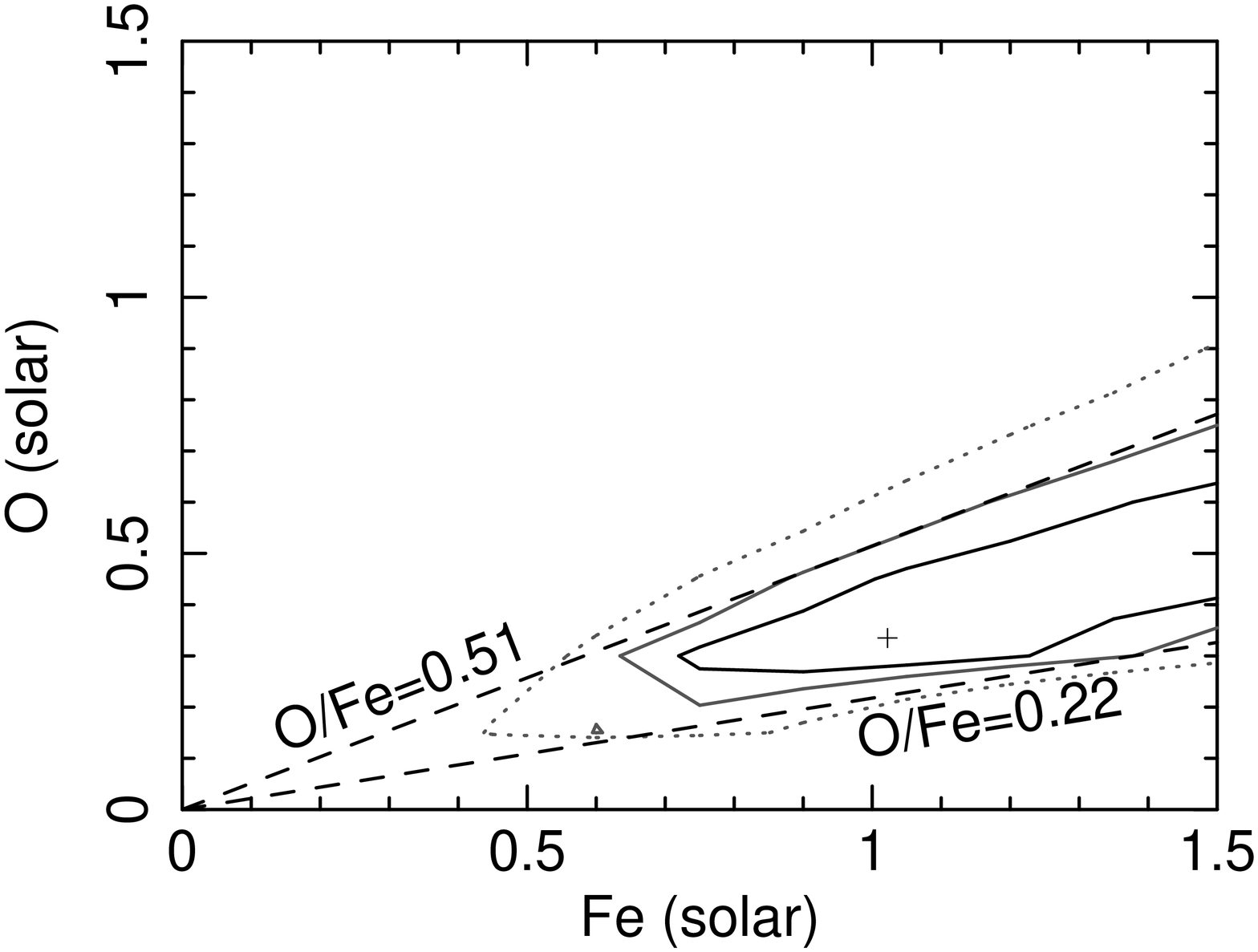}
    \FigureFile(80mm,80mm){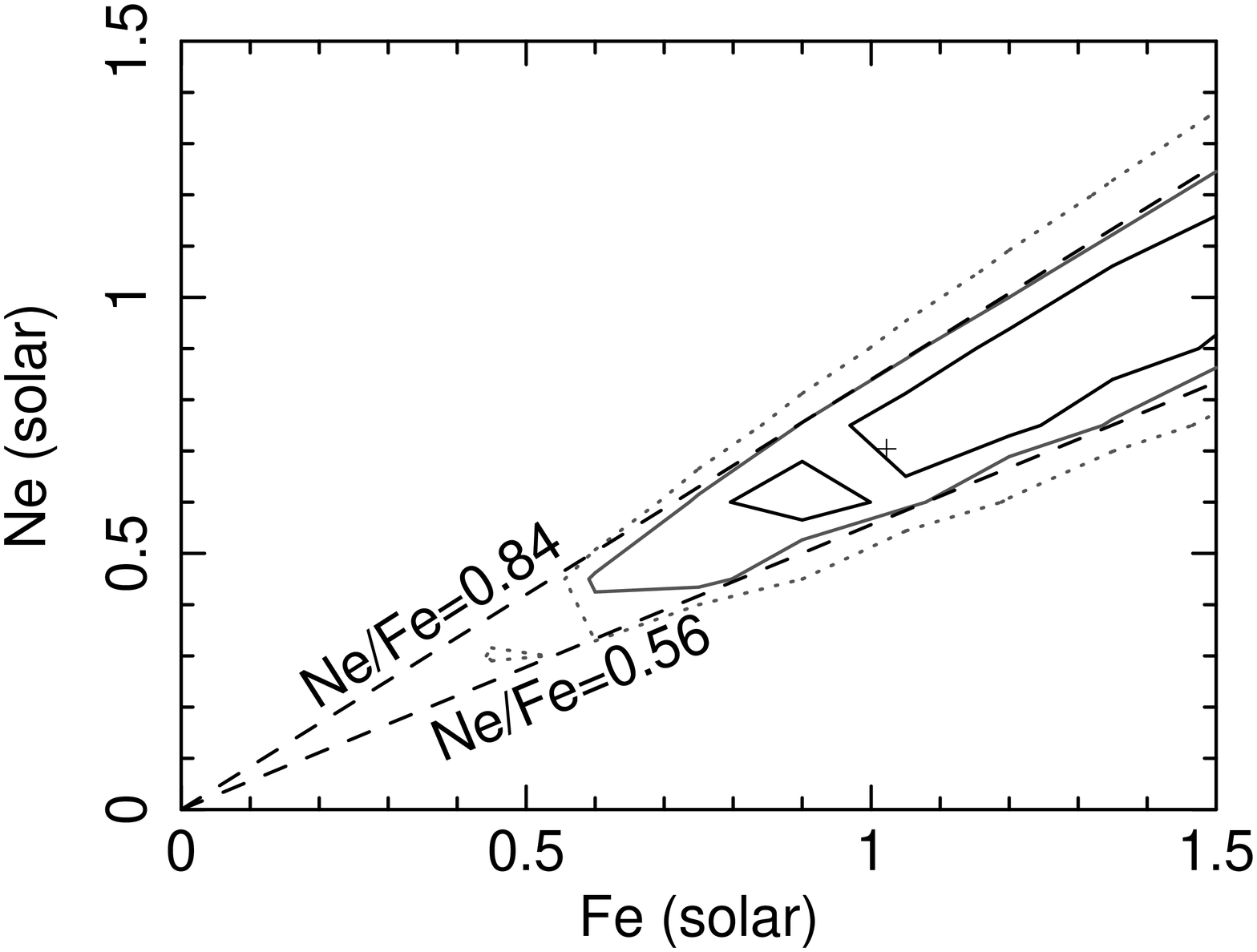}
    \FigureFile(80mm,80mm){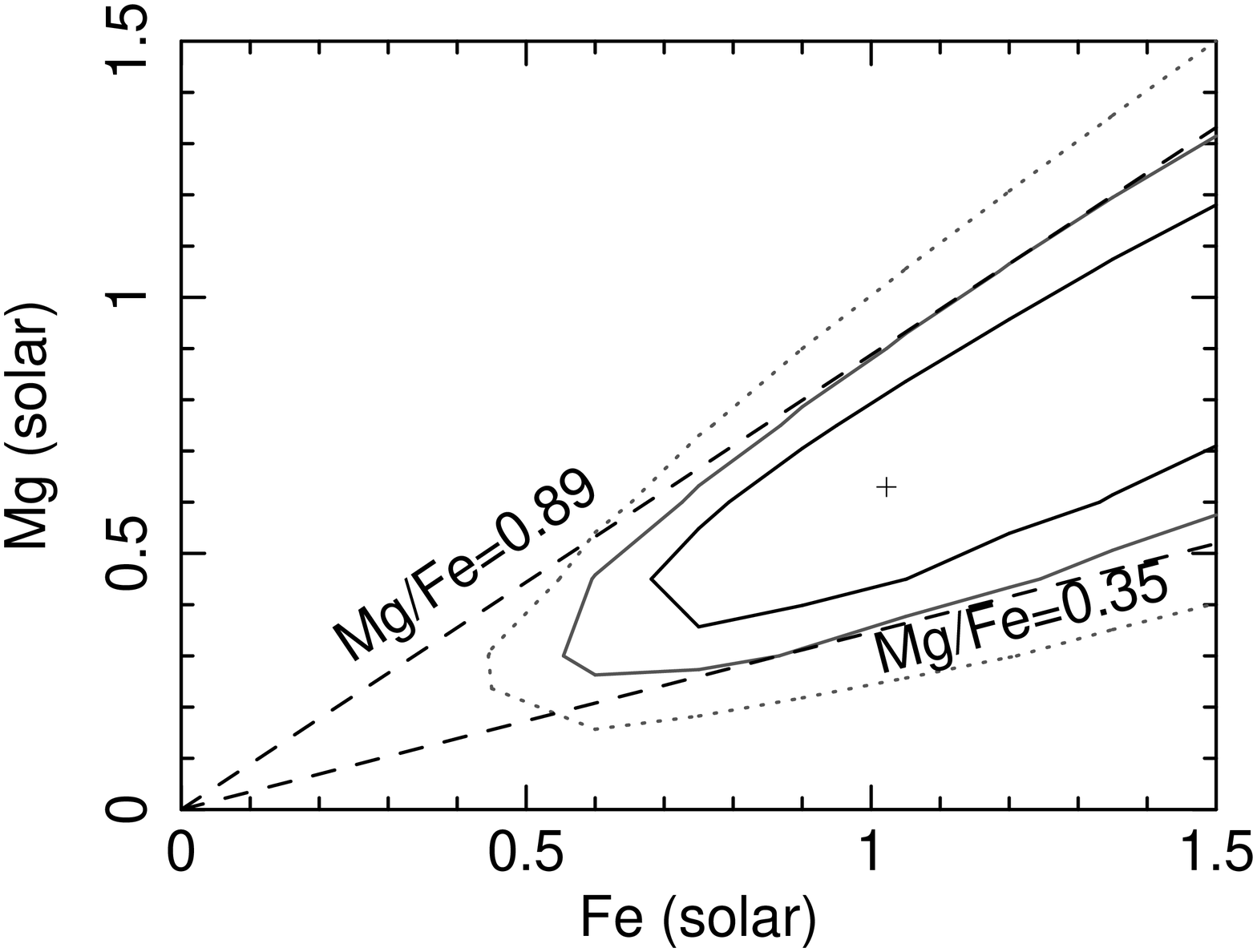}
  \end{center}
  \caption{Confidence contours for O, Ne, and Mg abundances vs. Fe
 abundance. Crosses denote the best-fit, and
 black-solid, gray-solid, and gray-dotted contours represent 68\%, 90\%, and 99\% confidence levels,
 respectively. Dashed lines correspond to
constant abundance ratios.}\label{fig:cont}
\end{figure*}

Figure \ref{fig:cont} shows the confidence contours of O, Ne, and Mg
abundances against Fe abundance. Fe in the ISM
originates from  type Ia and type II SNe, whereas, O, Ne, and Mg come 
from  type II SNe \citep{Iwa1999,Nomo2006}. Thus, O-, Ne-, and Mg-to-Fe abundance
ratios are particularly important for constraining the contributions from
type Ia and II SNe. These contours show elongated
shapes, especially for the Ne and O abundances vs. Fe abundance, and are along the lines
representing constant O/Fe, Ne/Fe, and Mg/Fe ratios. 
Thus, we can  constrain the metal-to-Fe abundance ratios well, 
although the individual error bars for each metal abundance are large. 

O/Fe, Ne/Fe and Mg/Fe  ratios derived from these
confidence contours are also summarized in Table \ref{tab:fit}.
Abundance ratios of O/Fe, Ne/Fe, and Mg/Fe in solar units
are $0.33^{+0.19}_{-0.11}$, $0.69^{+0.15}_{-0.13}$, and
$0.62^{+0.27}_{-0.27}$, respectively.  O abundance is
especially small relative to the Fe abundance. 

\subsection{Uncertainties in spectral fit}

In this section, we examine the uncertainties in the abundance
determination because of an uncertainty in the spectral modeling.

To investigate the systematic effect of the model for
unresolved sources, described in Section \ref{sec:us},
we also fitted the components with a bremsstrahlung model
(e.g., \cite{Matsu1994,Irwin2000}) and compared
the results with those using the power-law model.
Here, we fixed the temperature of the bremsstrahlung model  
 at 7.0 keV (\cite{Matsu1994}). The best-fit spectra are shown in
Figure \ref{fig:spec}. As shown in Table \ref{tab:fit},
the derived results are almost the same.

We also tried a spectral fitting with a two-temperature model for the ISM
(hereafter, the 2T model). However, the values of reduced-$\chi^2$ did not
improve, and the same residuals remained around 0.8 keV, at the Fe-L line energies.
The temperature of the first component was almost the same as that in
the 1T model, and the normalization of the second component was almost zero.
Thus, the derived abundances were unchanged.
We then fitted the spectra  with a five-temperature model
for the ISM (hereafter the 5T model); we fixed the ISM temperatures at 
(i) 0.2, 0.4, 0.6, 0.8, and 1.0 keV and (ii)  0.1, 0.2, 0.3, 0.4, and 0.5 keV.

\begin{figure*}
  \begin{center}
    \FigureFile(160mm,80mm){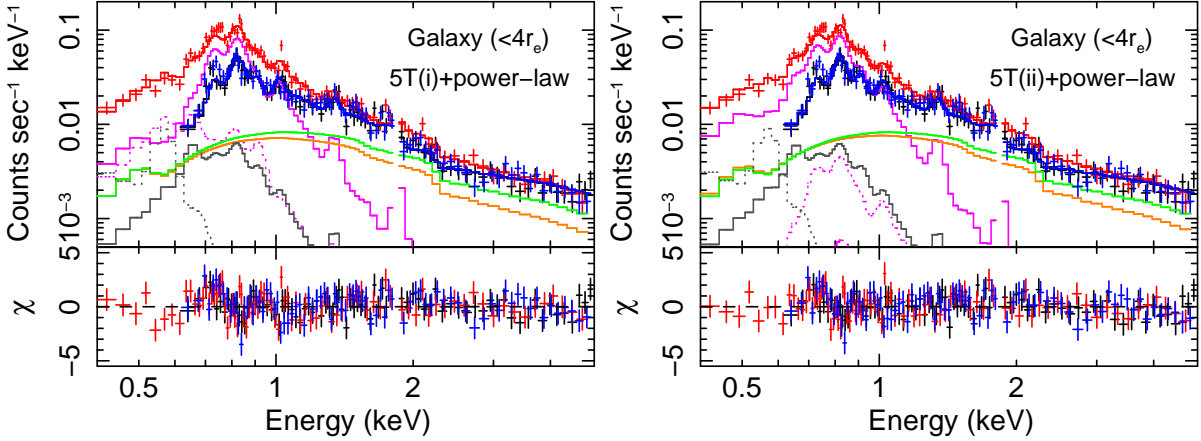}
  \end{center}
 \caption{Spectra within 4~$r_e$ observed
 with XIS0 (black), XIS1 (red), and XIS3 (blue). 
 These spectra are fitted with 5T(i)+power-law (left), and 5T(ii)+power-law (right) models.
 Lower panels show fit residuals.
 For simplicity, only model components for XIS1 spectra are
 shown. Colors have the same meanings as in Figure \ref{fig:spec}}\label{fig:spec5t}
\end{figure*}

Figure \ref{fig:spec5t} shows spectra fitted with the 5T(i) and
5T(ii) models. The spectral fitting with the 5T(i) model
provided a worse $\chi^2$ value than that with the 1T model
(Table \ref{tab:fit5t}).
Only the normalizations of  ISM
components with temperatures of 0.2 and 0.4 keV are $>$0, and the
normalizations of the other ISM components fall to 0. This suggests
that temperature components above 0.6 keV do not exist in the ISM of
NGC 4382. On the other hand, the spectral fitting with the 5T(ii) model
provides the same $\chi^2$ as the 1T model (Table \ref{tab:fit5t}).
In this fitting, only the normalizations of temperature components
at 0.3, 0.4, and 0.5 keV are $>$ 0 (cf. Figure \ref{fig:spec5t}).
The contribution of the 0.3 keV component is the largest of the three
temperature components at about 97\%. The derived abundances and abundance ratios did
not differ from the results of 1T spectral fitting. 
These 5T models even gave similar residual
structures around 0.8 keV. These discrepancies in the Fe-L energy range
are also seen in  Suzaku observations of NGC 720, NGC 1404, and
NGC 4636, whose ISM temperatures are also $\sim0.6$ keV
\citep{Matsu2007b,Tawara2008,Haya2009}. Therefore, these residual
structures are likely to be related to poorly modeled Fe-L lines.

To investigate the effect of the temperatures of the Galactic components
on the ISM abundances of NGC 4382, we allowed the temperature
of the MWH component to vary and re-fitted the spectra.
We got almost the same ISM temperature and abundances.

\begin{table*}
\begin{center}
\caption{Results of spectral fittings within 4~$r_e$ of NGC 4382 with 5T models}
\label{tab:fit5t}
\begin{tabular}{lcccccc}
\hline \hline
model & kT   & O       & Ne & Mg & Fe & $\chi^2$/d.o.f. \\
      & (keV) & (solar) & (solar) & (solar) & (solar) &                 \\ \hline
5T(i)+power-law & 0.2, 0.4, 0.6, 0.8, 1.0$^*$ & $  0.39 _{-  0.15 }^{+  0.22 } $ & $  0.68 _{-  0.32 }^{+  1.27 } $ & $  0.57 _{-  0.29 }^{+  1.25 } $ & $  0.97 _{-  0.40 }^{+  1.88 } $ &   783 /   655 \\
5T(ii)+power-law & 0.1, 0.2, 0.3, 0.4, 0.5$^*$ & $  0.32 _{-  0.12 }^{+  0.70 } $ & $  0.74 _{-  0.25 }^{+  0.89 } $ & $  0.69 _{-  0.28 }^{+  1.51 } $ & $  1.10 _{-  0.35 }^{+  2.24 } $ &   759 /   655 \\
\hline
\multicolumn{7}{l}{$^*$The values were fixed}\\
\end{tabular}
\end{center}
\begin{center}
\begin{tabular}{lccc}
\hline \hline
model & O/Fe & Ne/Fe & Mg/Fe \\
      & (solar) & (solar) & (solar) \\\hline
5T(i)+power-law & $  0.40 _{-  0.22 }^{+  0.38 } $ & $  0.70 _{-  0.26 }^{+  0.19 } $ & $  0.58 _{-  0.25 }^{+  0.26 } $ \\
5T(ii)+power-law & $  0.30 _{-  0.08 }^{+  0.15 } $ & $  0.67 _{-  0.14 }^{+  0.14 } $ & $  0.63 _{-  0.26 }^{+  0.27 } $ \\
\hline
\end{tabular}
\end{center}
\end{table*}

\subsection{Radial profiles}
\label{sec:radp}

\begin{figure}
  \begin{center}
    \FigureFile(80mm,80mm){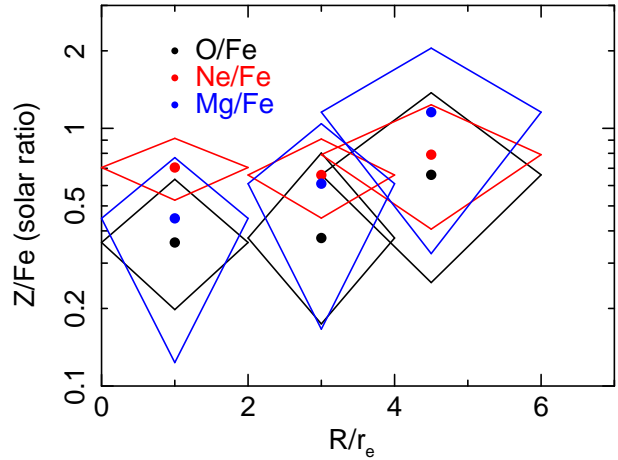}
  \end{center}
  \caption{Radial profiles of the abundance ratios of O,
 Ne, and Mg relative to the Fe abundance.}\label{fig:abratio_prof}
\end{figure}

To study radial discrimination, we
investigated the spectra extracted from the different radial regions,
$R < 2 r_e$, $2 r_e < R < 4 r_e$, and $3
r_e < R < 6 r_e$. Here, $R$ is the projected radius from the galaxy's center.
In Table \ref{tab:result_prof}, we summarized the fitting results.
A low signal-to-noise ratio does not constrain the abundances.
We
can derive the abundance ratios against Fe only, although each value
has large error bars.
Figure \ref{fig:abratio_prof} shows the derived radial profiles of 
O/Fe, Ne/Fe, and Mg/Fe ratios. 
On the whole, the derived abundance ratios of the O/Fe, Ne/Fe, and Mg/Fe
are consistent with those within 4~$r_e$, although these ratios have
large error bars.
At Suzaku's angular resolution, about 30\% of
the photons from the luminous central region within 2~$r_e$ escape into the ring of
2--4~$r_e$.  As a result, the derived abundance ratios of the  two
regions  are consistent with each other.
At 3--6~$r_e$, the signal-to-noise ratio is not high enough to constrain the
abundance pattern.

\begin{table*}[htbp]
\begin{center}
\caption{Radial results of  NGC 4382 spectral fittings}
\label{tab:result_prof}
\begin{tabular}{cccccccc}
\hline \hline
ring & model & kT   & O       & Ne & Mg & Fe & $\chi^2$/d.o.f. \\
     &       & (keV) & (solar) & (solar)   & (solar)   & (solar)   &                 \\ \hline
0--2~$r_e$ & 1T+power-law & $ 0.318 _{- 0.026 }^{+ 0.030 } $ & $  >0.17 $ & $  >0.39 $ & $  >0.17 $ & $  >0.60 $ &   638 /   532 \\
2--4~$r_e$ & 1T+power-law & $ 0.312 _{- 0.033 }^{+ 0.024 } $ & $ >0.15  $ & $  >0.30 $ & $  >0.41 $ & $  >0.49 $ &   631 /   538 \\
3--6~$r_e$ & 1T+power-law & $ 0.327 _{- 0.039 }^{+ 0.044 } $ & $  0.30 _{-  0.19 }^{+  3.13 } $ &
$  >0.12 $ & $  0.52 _{-  0.38 }^{+  5.45 } $ & $  0.45 _{-  0.26 }^{+ 10.46 } $ &   752 /   619 \\
\hline
\end{tabular}
\end{center}
\begin{center}
\begin{tabular}{ccccc}
\hline \hline
ring & model & O/Fe    & Ne/Fe   & Mg/Fe   \\
     &       & (solar) & (solar) & (solar) \\ \hline
0--2~$r_e$ & 1T+power-law & $  0.36 _{-  0.16 }^{+  0.27 } $ & $  0.70 _{-  0.18 }^{+  0.21 } $ & $  0.45 _{-  0.32 }^{+  0.32 } $ \\
2--4~$r_e$ & 1T+power-law & $  0.38 _{-  0.20 }^{+  0.43 } $ & $  0.66 _{-  0.21 }^{+  0.25 } $ & $  0.61 _{-  0.44 }^{+  0.43 } $ \\
3--6~$r_e$ & 1T+power-law & $  0.66 _{-  0.41 }^{+  0.71 } $ & $  0.79 _{-  0.38 }^{+  0.44 } $ & $  1.16 _{-  0.83 }^{+  0.89 } $ \\
\hline
\end{tabular}
\end{center}
\end{table*}

\section{Discussion}

\subsection{Comparison with abundance patterns in the ISM of elliptical galaxies}

\begin{figure*}
  \begin{center}
    \FigureFile(120mm,120mm){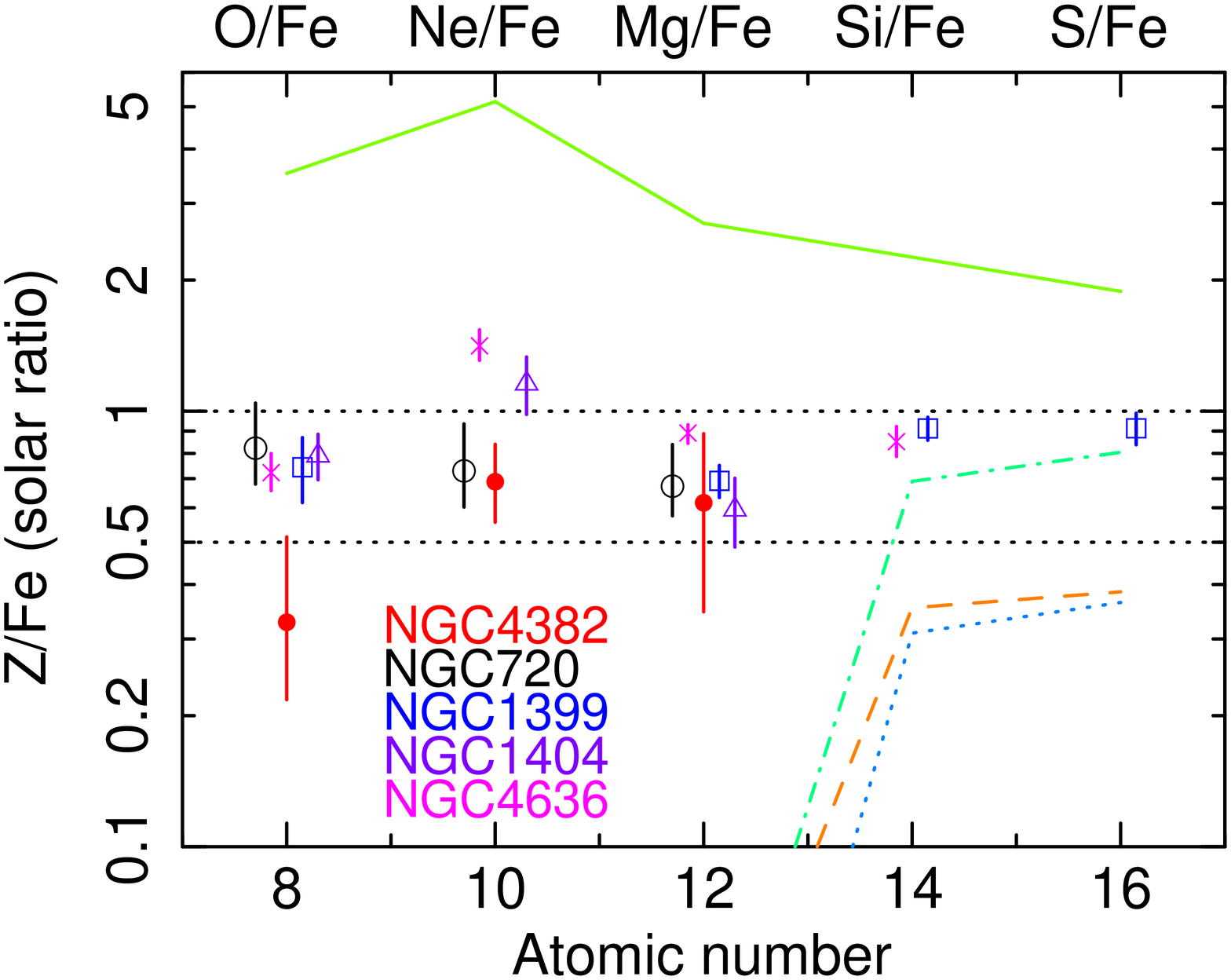}
  \end{center}
  \caption{Abundance ratios of O, Ne, and Mg against Fe in the ISM
of  NGC 4382 from the 1T model fit (filled red circles). For comparison, we plotted the abundance
 ratios of elliptical galaxies NGC 720 (black open circles; \cite{Tawara2008}), NGC
 1399 (blue open squares; \cite{Matsu2007b}), NGC 1404 (purple open triangles; \cite{Matsu2007b}),
 and NGC 4636 (magenta crosses; \cite{Haya2009}).
 Solid line represents the abundance pattern of SNe II \citep{Nomo2006},
 and dot-dashed, dotted, and dashed lines represent those of WDD1, WDD3,
 and W7 of SNe Ia, respectively \citep{Iwa1999}.}\label{fig:aratio}
\end{figure*}

Figure \ref{fig:aratio} shows the pattern of abundance ratios in the ISM
of  NGC 4382.
Here, we use those  derived from spectral
fitting with the 1T model (Table \ref{tab:fit}). 
The abundance ratios of O/Fe, Ne/Fe, and Mg/Fe have subsolar values
of $\sim$0.2--0.9 in solar units.  
O/Fe abundance ratio is obviously smaller than those of Ne/Fe
and Mg/Fe, although the values have large error bars.

Figure \ref{fig:aratio} also  compares the derived abundance pattern of NGC
4382 with those of other elliptical galaxies, NGC 720
\citep{Tawara2008}, NGC 1399 and NGC 1404 \citep{Matsu2007b}, and NGC
4636 \citep{Haya2009}. Here, the abundance values in solar units 
are converted using the new solar abundance table of \citet{Lodd2003},
since the other papers use different solar abundance tables, those of
\citet{Feld1992} or \citet{Ande1989}.
O/Fe and Ne/Fe abundance ratios increased by 60\%  than 
those  adopting the  solar abundances of
\citet{Feld1992}, while the Mg/Fe abundance ratio 
decreased by 30\%, compared with that adopting the solar abundances of
\citet{Ande1989}.

The abundance ratios of the four elliptical galaxies have a similar pattern with
small scatter, except for the Ne/Fe ratios, and are not very different from the
solar abundances of \citet{Lodd2003}.
The average ratios of O/Fe, Ne/Fe, and Mg/Fe for  these four elliptical
galaxies are 0.77, 1.10, and 0.71 in solar units, respectively.
As shown in Figure \ref{fig:aratio},
the O/Fe ratio of NGC 4382 is  smaller than those of the
elliptical galaxies. 
Adopting the 99\% confidence error, the upper O/Fe value of NGC 4382 becomes
 0.61 in solar units, which is still smaller than the values of the ellipticals.
The Ne/Fe ratio of the galaxy is consistent with the smallest value for 
the ellipticals.
The uncertainty in the spectral modeling of the Fe-L lines may cause
the scatter in  the Ne/Fe ratio in these galaxies and also in
the ICM in groups and clusters of galaxies (e.g. \cite{Matsu2001},
\cite{Komi2009}).
The Mg/Fe ratio of the S0 galaxy has a large error and 
is consistent with those of the  ellipticals.
Since  O is much more abundant  in number relative to H than Ne and Mg,
hereafter  we will discuss possible differences in the metal
enrichment histories of the ISM in NGC 4382 and the elliptical
galaxies.

\subsection{Contributions from SNe Ia and SNe II}

The theoretical SNe Ia and SNe II yields are also plotted in Figure
\ref{fig:aratio}. Here, we refer to the SNe Ia yields of the
W7, WDD1, and WDD3 models calculated by \citet{Iwa1999}.
The SNe II yields were taken from
\citet{Nomo2006}, which is an average over the Salpeter initial mass
function of stellar masses from 10 to 50 $M_{\odot}$ with a progenitor
metallicity of Z = 0.02.

The ISM abundance ratios of NGC 4382 and the elliptical galaxies are located
between those of SNe Ia and SNe II. This means that the metals in the ISM
are a mixture of SNe Ia and SNe II yields.

Since O is predominantly synthesized by SNe II,
the observed lower O/Fe ratio in the ISM of NGC 4382 indicates larger
contributions from SNe Ia to the ISM in this S0 galaxy than in
 elliptical galaxies. In early-type galaxies,
the ISM comes from stellar mass loss and is polluted by present SNe Ia.
To explain the low O/Fe ratio in NGC 4382, 
the present SN Ia rate may differ between NGC 4382 and
elliptical galaxies, or stars in this galaxy may contain more SNe Ia products
than those in elliptical galaxies.

\subsection{SN Ia rate of S0 and elliptical galaxies}

Assuming that the Fe abundance in the ISM is mainly the sum of stellar metallicity and
the contribution from SNe Ia, it can be expressed as 
\begin{eqnarray}
\left(\frac{\rm Fe}{\rm H}\right)_{\rm ISM}=\left(\frac{\rm Fe}{\rm H}\right)_{\ast}+\left(\frac{\theta_{\rm SN}M^{\rm Fe}_{\rm SN}}{\alpha_{\ast}}\right)(z^{\rm Fe}_{\rm solar})^{-1}\nonumber\\
=\left(\frac{\rm Fe}{\rm
  H}\right)_{\ast}+3.5\frac{\left(\frac{\theta_{\rm
  SN}}{1.0\times10^{-13}/{\rm yr}/L_B}\right)\left(\frac{M_{\rm SN}^{\rm
  Fe}}{0.6M_{\odot}}\right)}{\left(\frac{\alpha_{\ast}}{1.5\times
  10^{-11} M_{\odot}/{\rm yr}/L_B}\right)}~~~({\rm solar})
\end{eqnarray}
(see \cite{Matsu2003} for details). Here, (Fe/H)$_{\ast}$ is the stellar Fe
abundance synthesized by SNe Ia, $\theta_{\rm SN}$ is the SN Ia rate,
$M^{\rm Fe}_{\rm SN}$ is the mass of
Fe synthesized by one SN Ia, $\alpha_{\ast}$ is the mass loss rate of
stars, and $z^{\rm Fe}_{\rm solar}$ is the Fe metallicity of gas with
the solar abundance. Assuming the age to be 13 Gyr, 
we used the mass loss rate from
\citet{Ciot1991}, which is approximated by $1.5\times10^{-11}L_B
t_{15}^{-1.3}M_{\odot}/{\rm yr}$. Here $t_{15}$ is the age in units of 15 Gyr,
and $L_B$ is the B-band luminosity. 
$M_{\rm Fe}$ produced by one SN Ia
explosion is likely to be $\sim$ 0.6$M_{\odot}$ \citep{Iwa1999}.
Using the SN Ia rate in early-type galaxies, which was estimated to be $\sim$0.1--0.3
SN Ia/100 yr/$10^{10} L_{\odot}$ in previous studies
\citep{Cran1977,Barb1978,Capp1997,Capp1999,Tura1999,Shar2007,Mann2008},
the present contribution of SNe Ia to the Fe abundance in early-type galaxies
becomes 2.9--8.7 solar.

Recently, \citet{Mann2008}
reported that the cluster membership and morphology of early-type
galaxies may affect the SN rate. 
In clusters, the derived SN Ia rate of S0 galaxies,
$0.121^{+0.097}_{-0.059}$ SN Ia/100 yr/$10^{10}M_{\odot}$,
tends to be higher than the $0.053^{+0.029}_{-0.020}$ SN
Ia/100yr/$10^{10}M_{\odot}$ of elliptical galaxies.
Since the confidence level of these values is
1 $\sigma$, the difference is rather marginal.
Assuming the stellar B-band mass-to-light ratio of NGC 4382 and
the elliptical galaxies to be 3.3 and 5.0 $M_{\odot}/L_{\odot}$
\citep{Nagi2009}, respectively, the contributions to Fe abundance from SNe Ia in NGC
4382 and the elliptical galaxies are 
5.9--20.9 and 4.8--11.9 solar, respectively.
As a result of the
 large error bars, these values are consistent with each other.

The  Fe abundances expected from the observed SN Ia rate
are higher than the Fe abundances in the ISM in
the early-type galaxies observed with Suzaku.
This result indicates a low SN Ia rate, although the absolute values of
Fe abundances have rather large errors owing to the elongated shape of
the confidence contours.
We note that a higher SN rate might be related to younger stellar age,
in which case the stellar mass loss rate might also be higher.
For example, assuming a stellar age of 10 Gyr, the mass loss rate
increases by 1.4 times, and the contribution of SNe Ia to the Fe abundance 
decreases by 0.7 times.

\subsection{Stellar O/Fe abundance ratio and evolution of S0 galaxies}

The other possibility is that the lower O/Fe ratio of NGC 4382 
may be because of  a difference in the SN Ia products included in stars in this and
the elliptical galaxies. 
Since a longer star-formation time scale yields more SN Ia products in 
stars, the difference in  O/Fe ratio in stars can constrain the star-formation history.
Suzaku observes the ISM of the entire galaxy and constrains the abundance pattern
in stars of the entire galaxy, while optical observations are limited to
the very center.

Optical observations found higher abundances of Fe
than of $\alpha$-elements in stars in the central region (within 2
arcsec) of NGC 4382;
$[\alpha/{\rm Fe}] = 0.12\pm0.06$ at the 1 $\sigma$ confidence level \citep{McD2006}.
Generally, giant elliptical galaxies have higher stellar abundances
and higher $\alpha/{\rm Fe}$ ratios.
For example, the $[\alpha/{\rm Fe}]$ of NGC 720 is $0.37\pm0.05$
within  $r_e$/8 from the center of the galaxy \citep{Hump2006}.
Therefore, the low O/Fe ratio in the ISM of NGC
4382 may reflect a lower $[\alpha/{\rm Fe}]$ in stars.
XMM-Newton and Chandra observations show that about half of the photons from
NGC 4832 within 8~$r_e$ are emitted from the region within 2~$r_e$
\citep{Nagi2009}. 
Therefore, we can conclude that at least within a few $r_e$
the ISM shows the low O/Fe ratio. 
This region is much larger than that  within 2 arcsec, to which
the optical observations \citep{McD2006} are limited.
In other words, the Suzaku observations 
provide evidence that
 not only the stars in the central region of NGC 4382,
but also those within at least a few
$r_e$ have a smaller $\alpha/{\rm Fe}$
ratio than those of the four elliptical galaxies.

A number of authors found a clear correlation between the
$\alpha$/Fe ratio and system mass (e.g., \cite{Wort1992,Nela2005,Thom2005,Grav2007}).
\citet{Thom2005} found that galaxies with stellar velocity
dispersion $\sigma >$ 200 km ${\rm s^{-1}}$ have a relatively higher
[$\alpha$/Fe] ($>$ 0.2) and also have higher metallicities.
NGC 4382 has a lower ISM temperature of $\sim$0.3 keV,
and its stellar velocity dispersion $\sigma$ is 179$\pm$6 km ${\rm s^{-1}}$ \citep{McD2006}.
Therefore, the lower O/Fe ratio in NGC 4382 may reflect 
a correlation with the system mass.
Usually, these metallicity and $\alpha$/Fe correlations with system mass 
are  interpreted as showing that higher-mass systems 
have experienced more rapid star formation than lower-mass
galaxies (e.g. \cite{Thom2005,McD2006}). Thus, 
most stars in larger systems would have formed before SNe Ia began to synthesize
large amounts of Fe.

Considering the fractional evolution of S0 and spiral galaxies in
clusters, spiral galaxies might have changed into
S0 galaxies (e.g.
\cite{Dres1997,Fasa2000,Treu2003,Koda2004,Post2005,Smith2005,Desai2007,Pogg2009}). 
If so, stars in S0 galaxies would contain more SN Ia products than those
in elliptical galaxies,
reflecting the fact that stars in spiral galaxies have lower
$[\alpha/{\rm Fe}]$ ratios than those in giant ellipticals.
For example, the value of $[\alpha/{\rm Fe}]$ of the Sun is zero.
If S0 galaxies in clusters generally have lower O/Fe ratios in the ISM,
the hypothesis of morphology changes would be supported.
 To investigate further, we need more sample galaxies for which to derive
abundance patterns in the ISM using Suzaku observations.

\section{Summary and conclusion}

With the Suzaku observation of S0 galaxy NGC 4382, we measured the ISM
temperature, metal abundances of  O, Ne, Mg and Fe, and their abundance
ratios for the region within 4~$r_e$ of the galaxy's center. 
The temperature, 0.3 keV, and  O/Fe ratio, 0.3 in solar units, 
in the ISM of this galaxy are  smaller than
those in elliptical galaxies 
observed with Suzaku, NGC 720 \citep{Tawara2008}, NGC 1399 and
NGC 1404 \citep{Matsu2007b}, and NGC 4636 \citep{Haya2009}.
The lower O/Fe ratio of the ISM may reflect
a higher rate of present SN Ia or a lower $\alpha/{\rm Fe}$ ratio
in stars in  NGC 4382 \citep{McD2006}
than in those in the four elliptical galaxies.

\subsection*{}

We thank to the  referee for a careful reviewing and many helpful comments.
We also thank all members of the Suzaku hardware and software teams and
the science working group.

\end{document}